\documentstyle[12pt,epsfig]{article}
\begin{document}

\rightline{\large Feb 2003}

\vskip 2.3cm

\centerline{\Large Some comments on Super-Kamiokande's multi-ring
analysis }
\vskip 2.4cm
\centerline{\large R. Foot}
\vskip 0.7cm
\centerline{foot@physics.unimelb.edu.au}
\centerline{\large \it School of Physics}
\centerline{\large \it University of Melbourne}
\centerline{\large \it Victoria 3010 Australia}

\vskip 3cm
\noindent
The super-Kamiokande collaboration have used multi-ring events
to discriminate between
the $\nu_\mu \to \nu_\tau$ and $\nu_\mu \to \nu_s$ solutions
to the atmospheric neutrino anomaly. We show that the
effect of systematic uncertainties in cross sections are so significant
that the usefulness of multi-ring data to distinguish between
these two solutions is doubtful.

\newpage
There are two quite different interpretations of the
neutrino physics data.
One possibility is that
the solar and atmospheric neutrino anomalies result
from three flavour oscillations of the three known
neutrinos with two large mixing angles\cite{bm}.
While this scheme provides a very good fit to the
solar and atmospheric data, it cannot accommodate
the LSND data\cite{lsnd}. In fact, this three-flavour interpretation
of the atmospheric and solar neutrino anomalies
is disfavoured at the 4-5 sigma level from
the LSND neutrino experiment.

An alternative possibility invokes a fourth effectively
sterile neutrino ($\nu_s$)\cite{other}, which allows each neutrino physics
anomaly to be explained by 2-flavour oscillations\cite{fv}.
[Solar by large angle $\nu_e \to \nu_\tau$ oscillations,
atmospheric by large angle or maximal $\nu_\mu \to \nu_s$
oscillations
and LSND by $\bar \nu_e \to \bar \nu_\mu$ oscillations].
In this case the LSND data can be explained,
however a subset of the super-Kamiokande
data disfavour this possibility at about the 1.5-3 sigma
level (depending on how the data are analysed)\cite{sk2,f}.


Both schemes provide an acceptable global fit
of the data\footnote{The global goodness of fit (gof)
of the above 4 neutrino scheme was explicitly calculated in Ref.\cite{f2}
to be 0.27, which means that there is a 27\% probability of obtaining
a worse global fit to the data.}, although each scheme fails to provide
a good fit to some particular data subset.
In the case of the 3 neutrino scheme the LSND data is not
explained, while in the case of the above 4 neutrino model
a subset of super-Kamiokande data are poorly fitted.

Given this current situation, it makes sense to
a) check the LSND result, which is being done
by MiniBooNE\cite{boone} and b) check the $\nu_\mu \to \nu_\tau$
versus $\nu_\mu \to \nu_s$ discrimination.
The $\nu_\mu \to \nu_\tau$ versus $\nu_\mu \to \nu_s$
discrimination will eventually be done by the
long baseline neutrino experiments. In the meantime
it makes sense to examine the robustness of super-Kamiokande's
claim\cite{sk2} that the $\nu_\mu \to \nu_s$ hypothesis is disfavoured
by a subset of the super-Kamiokande data.

Super-Kamiokande identified three subsets
of data which might potentially discriminate between
the $\nu_\mu \to \nu_\tau$ and $\nu_\mu \to \nu_s$
solutions to the atmospheric neutrino anomaly\cite{sk2}.
However, this discrimination, was not particularly
stringent (super-Kamiokande found that the $\nu_\mu \to \nu_s$
possibility had a goodness of fit (gof) of about $0.01$), and an alternative
analysis\cite{f} of
the same data (but using binned data
rather than ratios) gave a much better fit (a
gof of about $0.1$).

One of the issues discussed in Ref.\cite{f}
was that of systematic uncertainties. Two
main sources of systematic uncertainties were
identified. In the through going muon data set,
one can have atmospheric muons masquerading as
neutrino induced muons in the (near) horizontal
bin. It is difficult to reliably estimate such
contamination, and if it is larger than expected then
the resulting neutrino induced through going muon zenith angle
distribution will be flatter, which improves the
fit of the $\nu_\mu \to \nu_s$ oscillation hypothesis.

In the multi-ring analysis, uncertainties
in the cross sections are expected to be significant\cite{f}.
The main purpose of this paper is to examine
the effect of cross sectional uncertainties in
the up-down multi-ring ratio used by super-Kamiokande.
As we will show, these uncertainties are quite large -- much larger
than those quoted by super-Kamiokande. In fact, our analysis suggests
that multi-ring data, at least with the super-Kamiokande cuts, cannot
provide a useful discrimination between the $\nu_\mu \to \nu_\tau$
and $\nu_\mu \to \nu_s$ hypothesis.


Super-Kamiokande employed the following selection criteria
for their multi-ring analysis:
(1) vertex within the fiducial volume and no exiting
track; (2) multiple Cerenkov rings; (3)
particle identification of the brightest ring is e-like;
(4) visible energy greater than 400 MeV.
According to the super-Kamiokande monticarlo\cite{sk2},
events satisfying this selection criteria
(assuming no oscillations) will be made up of
Neutral Current (NC), $\nu_e$ Charged Current ($\nu_e CC$),
and $\nu_\mu$ Charged Current ($\nu_\mu CC$) events in proportion:
\begin{eqnarray}
NC \ 29\%, \ \nu_e CC \ 46\%, \ \nu_\mu CC \ 25\%
\ \ \ \left[{\rm no \ oscillations}
\right]
\label{1}
\end{eqnarray}

For maximal $\nu_\mu \to \nu_s$ oscillations
with $\delta m^2 \sim 3\times 10^{-3}\ eV^2$
approximately half of
the up-going $\nu_\mu$ oscillate into $\nu_s$.
This will lead to an overall up-down asymmetry for
the multi-ring events due to $\nu_\mu$ interactions.
It is convenient to define the quantities, $r_1, r_2$ and
$R$ where
$r_1$ is the up-down ratio for $\nu_\mu CC$ events,
$r_2$ is the up-down ratio for $\nu_\mu$ induced NC events
and
$R$ is the
ratio of $\nu_e/(\nu_\mu + \nu_e)$ events in NC ($R \approx
0.25$).
With these definitions, the multi-ring up-down ratio
(with `up' defined as $\cos\theta < -0.4$, and `down' with
$\cos\theta > 0.4$, i.e. the same definition as used
by super-Kamiokande) is simply:
\begin{eqnarray}
r(\nu_\mu \to \nu_s) &=& {0.46 + 0.25*r_1 + 0.29*R + 0.29*(1-R)*r_2}
\nonumber \\
&\simeq & 0.53 + 0.25*r_1 + 0.22*r_2
\label{yyx}
\end{eqnarray}
If there were perfect angular correlation, we expect
$r_1 = r_2 = 0.5 $ for maximal oscillations.
In this extreme case, one expects [from Eq.(\ref{yyx})] $r \simeq 0.77$.
But the angular correlation is not perfect because
of the momentum of the recoiling nucleus is not observed. For NC events,
one has in addition the momentum from the scattered neutrino (and
there are also undetected neutrinos from decaying pions and charged
particles/photons below the Cerenkov threshold).
So one would
expect something like $r_1 \approx 0.55$, $r_2 \approx 0.60$ to
be more realistic, suggesting a $r(\nu_\mu \to \nu_s) \approx 0.80$.


In the $\nu_\mu \to \nu_\tau$ case, things are a little
different since all of the NC events are up-down symmetric and there
are additionally up-down asymmetric
$\nu_\tau CC$ events (about $3\%$ according to the
super-Kamiokande MC\cite{sk2}). Assuming perfect angular
correlation, so that the $\nu_\tau CC$ events
are all up-going,
the net effect is an up-down ratio of:
\begin{eqnarray}
r (\nu_\mu \to \nu_\tau) = 0.75 + 0.25*r_1 + 0.06
\label{3x}
\end{eqnarray}
Perfect angular correlation for the $\nu_\mu CC$ events would imply
$r_1 = 0.5$, giving $r(\nu_\mu \to \nu_\tau) \approx 0.93$,
Of course, the angular correlation is not perfect, with
$r_1 \approx 0.55$, but this should be roughly compensated by the
$\nu_\tau CC$ events
giving some down going events.
In summary, we estimate a multi-ring up-down asymmetry for
the $\nu_\mu \to \nu_s$ and $\nu_\mu \to \nu_\tau$ cases of:
\begin{eqnarray}
r(\nu_\mu \to \nu_s) &\simeq & 0.53 + 0.25*r_1 + 0.22*r_2 \simeq 0.80 \nonumber \\
r(\nu_\mu \to \nu_\tau) &\simeq & 0.75+ 0.25*r_1+ 0.06 \simeq 0.93
\end{eqnarray}
According to super-Kamiokande, figure 1b\cite{sk2}, their MC agrees
with the above estimates (within $1\%$ for
$\delta m^2 \sim 3 \times 10^{-3}\ eV^2$).

Note that there is only a small difference between the expected
$r(\nu_\mu \to \nu_s)$ and $r(\nu_\mu \to \nu_\tau)$ values (about
14\%), so the estimation of
systematic errors is extremely important.
Because the multi-ring sample, as defined by super-Kamiokande,
involves a combination of up-down symmetric events and
up-down asymmetric events, uncertainties in the relative proportion of these
events will lead to
systematic uncertainties in $r$.
The number of events satisfying the super-Kamiokande
selection criteria, coming from the three main processes,
NC, $\nu_e$ CC, $\nu_\mu$ CC have significant
uncertainties because of uncertainties in the size and shape
of the differential cross-sections involved.


We now try to estimate these
systematic uncertainties in the $r$ ratio due to these cross sectional
uncertainties.
Given the super-K selection criteria, the cross sectional rate uncertainties of the NC, $\nu_e CC$,
$\nu_\mu CC$, $\nu_\tau CC$ induced
multi-ring events should be approximately uncorrelated.
This is because the underlying processes are quite different,
as illustrated in Figure 1.
\vskip 0.5cm
\centerline{\epsfig{file=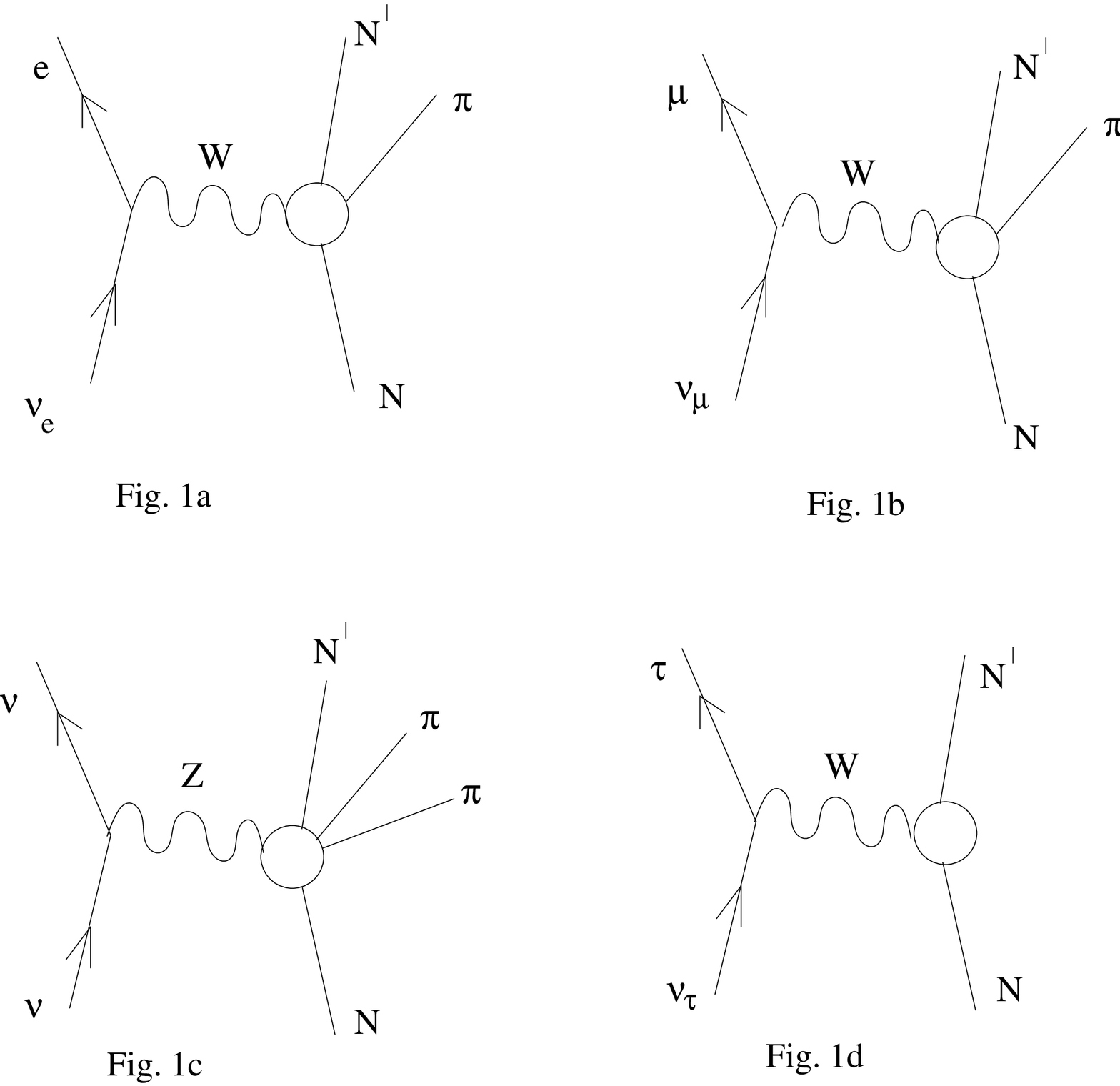,width=10.0cm}}
\vskip 0.5cm
\noindent
{\small Figure 1a,b,c,d: Examples of the four types of multi-ring
processes.
The large `O' represents non-perturbative strong interaction
physics.
Figure 1a: $\nu_e$ CC induced events.
The $\pi$ can be a $\pi^0$ or a charged
pion (the charge, depending on whether one has a neutrino
or anti-neutrino induced event). In this case the
the brightest e-like ring can originate from
the electron or the pion (if the pion is neutral).
Figure 1b: $\nu_\mu$ CC induced event. In this
case the emitted pion should be neutral, and the
e-like event must come from the photon(s) of the
decaying neutral pion. In this case the pion must
therefore be produced with an energy greater than the muon
to satisfy the super-Kamiokande selection
criteria (this makes this process quite different
from the process Fig.1a).
Figure 1c: NC events. In this case, one expects
two (or more) $\pi$ production to account for most of the
events. A single $\pi^0$ could lead to
two e-like rings, however a $\pi^0$
produced at high momentum the
rings merge cannot be distinguished.
Figure 1d: $\nu_\tau$ CC events. The decay of
the tauon can lead to a multi-ring event.}
\vskip 1cm
Note that the $\nu_e CC$ and $\nu_\mu CC$ events
appear very similar. Yet, the cross section
uncertainties are largely uncorrelated even
in this case. This is because of the super-Kamiokande
selection criteria for multi-ring events
which demands that the brightest ring is
e-like. This selection criteria strongly enhances
the $\nu_e CC$ process over the $\nu_\mu CC$ process,
since in the former case, the brightest e-like ring
typically comes from the emitted electron while in the
latter case it must come from the emitted pion.
Thus, in the $\nu_e CC$ process, the lepton is (typically) emitted with more
energy than the pion, while in the $\nu_\mu CC$ process, the
lepton is emitted with less energy than the pion.
Therefore, the cross sections for $\nu_e CC$ and $\nu_\mu CC$
events are {\it not} correlated.\footnote{
In fact, the integrated cross section for producing a $\nu_e CC$ events
is about 5 times larger than for a $\nu_\mu CC$
event. This can be seen from a) Eq.(\ref{1}), where
the number of
$\nu_e CC$ events is nearly double that of
the $\nu_\mu CC$ events ($46\%$ c.f. $25\%$)
and b) the flux of $\nu_e$ is
only about $1/3$ that of $\nu_\mu$.}

Define $N_{\nu_e CC}, N_{NC}, N_{\nu_\mu CC}, N_{\nu_\tau CC}$ to be the number
of $\nu_e CC$ events, NC events, $\nu_\mu CC$
events and $\nu_\tau CC$ events all satisfying the super-Kamiokande selection
criteria and assuming no oscillations. In this case, Eqs.(\ref{yyx},\ref{3x}),
can be equivalently expressed as:
\begin{eqnarray}
r(\nu_\mu \to \nu_s) &=& {N_{\nu_e CC} + N_{\nu_\mu CC} * r_1
+ R*N_{NC} + (1 - R)* N_{NC} * r_2
\over
N_{\nu_e CC} + N_{NC} + N_{\nu_\mu CC} }
\nonumber \\
r(\nu_\mu \to \nu_\tau) &=& {N_{\nu_e CC} + N_{NC} + N_{\nu_\mu CC} *
r_1 + 2N_{\nu_\tau CC} \over
N_{\nu_e CC} + N_{NC} + N_{\nu_\mu CC} }
\label{22}
\end{eqnarray}
The effect on the multi-ring up-down $r$
of uncorrelated cross section uncertainties in
the $\nu_e CC$, $\nu_\mu CC$, $NC$ and $\nu_\tau CC$ processes
can now be easily evaluated.
We denote the
percentage uncorrelated uncertainties
in $N_{NC}, N_{\nu_e CC},  N_{\nu_\mu CC}, N_{\nu_\tau CC}$
by:
\begin{eqnarray}
\delta_{NC}, \ \delta_{\nu_e CC}, \ \delta_{\nu_\mu CC}, \ \delta_{\nu_\tau
CC}
\end{eqnarray}
That is, the number of $\nu_e CC$ events is $N_{\nu_e CC} (1 \pm
\delta_{\nu_e CC})$, etc.
We now estimate the corresponding uncertainty in the Super-Kamiokande
multi-ring up-down ratio, $r$.
Let us take
the $\nu_e$ induced CC events as an example.
If the number of $\nu_e$ induced CC events is uncertain by an amount
$\delta_{\nu_e CC} N_{\nu_e CC}$ then
the corresponding uncertainty in $r$ is
\begin{eqnarray}
\delta r &\approx & {\partial r \over \partial N_{\nu_e CC}}
\delta N_{\nu_e CC} \nonumber \\
&\approx& {\partial r \over \partial N_{\nu_e CC}}
\delta_{\nu_e CC} N_{\nu_e CC}.
\end{eqnarray}
In the case of the $\nu_\mu \to \nu_s$ hypothesis,
from Eq.(\ref{22}), we have
\begin{eqnarray}
N_{\nu_e CC}{\partial r \over \partial N_{\nu_e CC}}
&=& \left[1 - r_0 (\nu_\mu \to \nu_s) \right]
N_{\nu_e CC}/(N_{\nu_e CC} + N_{NC} + N_{\nu_\mu CC})
\nonumber \\
&=& 0.46*\left[ 1 - r_0 (\nu_\mu \to \nu_s)
\right]
\end{eqnarray}
where $r_0 (\nu_\mu \to \nu_s)$ is the estimate of the `central' value of
$r$.
Putting $r_0 (\nu_\mu \to \nu_s) \approx 0.80$ leads
to an uncertainty in $r$ of $0.09\delta_{\nu_e CC}$, implying a percentage
uncertainty
in $r$ of $\delta r/r_0 \approx 0.11 \delta_{\nu_e CC}$. A similar
procedure for the other processes, gives the following uncertainty in $r$:
\begin{eqnarray}
r(\nu_\mu \to \nu_s) &=& r_0 \left( 1 \pm 0.04\delta_{NC}
\pm 0.11\delta_{\nu_e CC} \pm 0.08 \delta_{\nu_\mu CC}\right)
\nonumber \\
r(\nu_\mu \to \nu_\tau) &=& r_0 \left( 1 \pm 0.02\delta_{NC}
\pm 0.03\delta_{\nu_e CC} \pm 0.10 \delta_{\nu_\mu CC} \pm
0.06\delta_{\nu_\tau CC} \right)
\label{xx}
\end{eqnarray}
Now, the precise values for the $\delta_i$ needs careful consideration.
The multi-pion production cross sections at energies of $\sim GeV$ are
quite poorly measured with uncertainties in the range
$30-40\%$\cite{paolo}.
We will assume
a $35\%$ uncorrelated uncertainty in $N_{\nu_e CC}$,
$N_{\nu_\mu CC}$ and
$N_{NC}$.  This implies a total uncertainty in $r$ (adding the
various uncertainties in quadrature) of
about:
\footnote{
Of course, systematic uncertainties are not Gausian etc...
so the usual cautionary remarks are in order.}
\begin{eqnarray}
r(\nu_\mu \to \nu_s) &=& r_0 (\nu_\mu \to \nu_s) (1 \pm 0.05)
\nonumber \\
r(\nu_\mu \to \nu_\tau) &=& r_0 (\nu_\mu \to \nu_\tau) (1 \pm 0.04)
\label{8}
\end{eqnarray}

Thus, we estimate that the theoretical expectation for
the super-Kamiokande $r$ ratio for their multi-ring events
has systematic uncertainties of about 5\% (in the
$\nu_\mu \to \nu_s$ case).
The super-Kamiokande
claim\cite{sk2} that the uncertainty in $r$ due to the
cross section uncertainties
is less than 1\% is very hard to understand. Indeed to get a 1\% uncertainty
in $r$ would require
cross sectional uncertainties 5 times
smaller than their expected value. In other words, the uncertainties
in the NC, CC rates would have to be less than about 6\%, which
does not seem possible. At the very least, such low uncertainties would
have to be carefully and explicitly justified.

So far, we have tried to estimate the systematic uncertainties
in $r$ due to cross-sectional uncertainties. It would also be
useful to try and independently estimate the central
values, Eq.(\ref{1}). Indeed,
the proportion of NC events found by super-Kamiokande seems somewhat
larger than might be expected.
If there were only one pion produced, then the process in
figure 1b might be expected to be roughly
the same magnitude as the process in
figure 1c. But, this is not the case.
The problem is that the pion needs
to be quite relativistic ($E > 400$ MeV) for
its direction to be correlated enough with
the incident neutrino to have significant
up-down asymmetry. Yet, in that case, the two e-like rings
tend to merge. If they didn't one could simply select
events with 2 e-like rings with invariant mass of
$m_\pi$ (and with visible energy greater than 400 MeV) to
to obtain an almost pure NC sample. Such a sample
would have a up-down
asymmetry of $r_2 \approx 0.6$ for $\nu_\mu \to \nu_s$
and 1 for $\nu_\mu \to \nu_\tau$ and no annoying cross-section
uncertainties to worry about! This would allow
for a sensitive discrimination between $\nu_\mu \to \nu_\tau$
and $\nu_\mu \to \nu_s$ modes\cite{pak}.
Unfortunately this is not possible because a $\pi^0$ with
high momentum produces two photons in nearly the same direction
and the two rings merge into one\cite{learn}.
Hence to obtain 2 rings, one needs multi-pion production for
the NC process. One might therefore expect that the number of NC events
to be significantly less than the number of $\nu_\mu CC$
events.

In conclusion, we have critically examined the effect of systematic
cross section uncertainties
in super-Kamiokande's multi-ring analysis.
Our study indicates that the systematic error is significant --
approximately 5\% in the up-down ratio. This is important because
the difference between the $\nu_\mu \to \nu_s$ and $\nu_\mu \to \nu_\tau$
hypothesis is only 14\%. Unfortunately, it seems that the use of multi-ring
data
to convincingly discriminate between the $\nu_\mu \to \nu_\tau$ and
$\nu_\mu \to \nu_s$ hypothesises is doubtful.

\vskip 0.6cm
\noindent
{\bf Acknowledgements}
\vskip 0.3cm
\noindent
The author thanks J. Learned, P. Lipari and R. Volkas for
discussions and correspondence.

\end{document}